\documentclass[sigplan,screen,authorversion=true, nonacm]{acmart}
\usepackage{hyperref}
\usepackage[hyphenbreaks]{breakurl}

\usepackage[inline]{enumitem}

\setcopyright{acmcopyright}
\copyrightyear{2018}
\acmYear{2018}
\acmDOI{XXXXXXX.XXXXXXX}

\acmConference[Conference acronym 'XX]{Make sure to enter the correct
  conference title from your rights confirmation emai}{June 03--05,
  2018}{Woodstock, NY}
\acmPrice{15.00}
\acmISBN{978-1-4503-XXXX-X/18/06}

\usepackage{xspace}
\newcommand{\UniMath}{\textsf{UniMath}\xspace}
\newcommand{\Coq}{\textsf{Coq}\xspace}
\newcommand{\cfont}[1]{\mathsf{#1}}
\newcommand{\Id}[1]{\cfont{Id}}
\newcommand{\constfont}[1]{\ensuremath{\mathcal{#1}}}
\newcommand{\cat}[1]{\ensuremath{\constfont{#1}}\xspace}
\newcommand{\CC}{\cat{C}}
\newcommand{\A}{\cat{A}}

\begin{document}

\title{Formalizing Monoidal Categories and Actions for Syntax with Binders}

\author{Benedikt Ahrens}
\email{B.P.Ahrens@tudelft.nl}
\orcid{https://orcid.org/0000-0002-6786-4538}
\affiliation{%
  \institution{Delft University of Technology}
  \country{The Netherlands}
}
\affiliation{
  \institution{University of Birmingham}
  \country{United Kingdom}
}

\author{Ralph Matthes}
\email{ralph.matthes@irit.fr}
\orcid{https://orcid.org/0000-0002-7299-2411}
\affiliation{%
  \institution{IRIT, Université de Toulouse, CNRS, Toulouse INP, UT3}
  \city{Toulouse}
  \country{France}
}

\author{Kobe Wullaert}
\email{K.F.Wullaert@tudelft.nl}
\orcid{https://orcid.org/0000-0003-4281-2739}
\affiliation{%
  \institution{Delft University of Technology}
  \country{The Netherlands}
}

\begin{abstract}
  We discuss some aspects of our work on the mechanization of syntax and semantics in the \UniMath library, based on the proof assistant \Coq. We focus on experiences where \Coq (as a type-theoretic proof assistant with decidable type-checking) made us use more theory or helped us to see theory more clearly.
\end{abstract}

\maketitle

\section*{Introduction}

We will report on an effort conducted since 2015 to formalize
categorical semantics of syntax with binders in constructive type
theory. The formalization is a contribution to the \UniMath library \cite{UniMath},
which is a library of \Coq vernacular files that uses \Coq in a
deliberately restricted way but adds the univalence axiom and
deactivates the checking of universe levels so as to accommodate
Voevodsky's resizing axioms. Besides the univalence axiom, it is in
general informed by univalent mathematics, in particular as regards
the definition of propositions (as contractible types) and sets (as
types where all equality types are propositions).

This effort entered a new phase in 2021 when the link to more abstract
formulations of (signatures for) syntax in monoidal categories (such as by way of actions) was
explored and formalized and when categorical constructions were built
more modularly by using displayed categorical notions. Furthermore,
bicategories were used for two purposes:
firstly, as a means of organizing
families of categories (the usual purpose of bicategories),
and, secondly, 
for abstracting away from the concrete situation of functors and
natural transformations in functor categories, by referring instead to
an abstract bicategory that can then be instantiated to the bicategory
of categories, functors, and natural transformations.

In 2022, we identified a reformulation of the notion of monoidal
categories that is no longer based on a notion of tensor as a functor
from the product of its underlying category with itself. Instead, it uses a
crafted notion of \emph{whiskered bifunctor} that allows us to smoothly
introduce a displayed notion of monoidal category.

We will highlight
\begin{enumerate*}[label=({\arabic*})]
\item how the concept of monoidal category is beneficial to type-check formalization
  about syntax with binders in intensional type theory,
\item how the abstraction provided by bicategories simplified our formalization involving functors and natural transformations,
\item why it is better to have a hand-crafted notion of bifunctor playing the role of the tensor product in monoidal categories, and
\item what are our preliminary findings about the benefit of formalizing categorical recursion schemes as regards to having a minimal set of assumptions.
\end{enumerate*}
We briefly discuss these points below.

\section{The Need for Monoidal Categories}\label{sec:monoidal}
When we formalize commuting diagrams as equations in \Coq, the minimum
requirement is that the source (and the target) objects of the equated
morphisms agree, and this with respect to convertibility, i.\,e.,
detectable by the type-checker. When the morphisms happen to be
natural transformations between functors and functor composition is
involved in those source ``objects'', a mathematician will always
argue modulo associativity of composition and implicitly use the
identity as its neutral
element. However, \UniMath does not provide those latter laws as convertibility.
While
$\Id{}\cdot F\Rightarrow G$ and $F\Rightarrow G$ are convertible types
of natural transformations, the functors $\Id{}\cdot F$ and $F$ are
not convertible,\footnote{Be aware that in \UniMath, composition is written in diagrammatic order instead; details are found in~\url{https://github.com/UniMath/UniMath/blob/7145259b9cf4f4ba6342943f67736ddcfead9a48/UniMath/CategoryTheory/UnitorsAndAssociatorsForEndofunctors.v}} and, consequently, neither are $H(\Id{}\cdot F)$ and $HF$, for $H$ a functor on the functor category where $F$ lives.
This scenario played an important role
in previously published work of the first two authors \cite{DBLP:conf/types/AhrensM15}.

Our way out is to explicitly add polymorphic functions that are pointwise the identity morphism and that mediate between those functors. The
concept of monoidal categories gives a rationale for these insertions and abstracts away from the concrete situation of functors and natural transformations where those pointwise identities suffice. Thus, monoidal categories provide a clean (and much more general) mathematical basis for inserting some pointwise identities to make compositions well-typed.

However, a more interesting use of monoidal categories is
that monoids therein are an abstraction of monads---when the monoidal
category is canonically derived from an endofunctor category, the two
concepts coincide. This use of monoidal categories was advocated already in 1999 \cite{DBLP:conf/lics/FiorePT99}. A formalization of such ideas was given in \Coq as of 2020 \cite{lafontskewedmoncat} %
and further work is in progress.

\section{The Need for Bicategories}\label{sec:bicat}
Monoidal categories are nothing but one-object bicategories, but using
monoidal categories does not mean using the conceptualization offered
by bicategories. For a fixed category, its endofunctors
and their natural transformations constitute a monoidal
category---this is the application scenario of the previous
section. However, in previous work involving the first and second
author \cite{DBLP:conf/cpp/AhrensMM22} on categorical representation
of multi-sorted syntax with binding, some building blocks of the
considered functors are not endofunctors. Hence a categorical modeling
of all categories, functors and natural transformations seems
appropriate, and this led us to consider the bicategory of categories. For the
sake of this abstract, we focus on an element of feasibility, not on
organizing the semantic elements---most notably actions of monoidal
categories on categories---again into bicategories. Section 4.3 of the
cited work argues that proving complicated equations between chains of
composed natural transformations are ``incommensurate with their
proofs on paper, the main problem being that there is too much
structure in those concrete functors and natural transformations, so
that any attempt at simplifying the goals at hand makes them
unreadably convoluted''. Our way out there was to abstract the problem
and suitably replace functors by one-cells and natural transformations
by two-cells of a(n unspecified) bicategory, even if we are finally
interested only in the bicategory of categories as an instance.

\section{The Need for Hand-crafted Bifunctors}\label{sec:bifunctor}

There was an identified yet unresolved gap in the cited previous work
\cite[Section 4.3]{DBLP:conf/cpp/AhrensMM22}: parameterized
distributivity for actions could not be obtained from strong monoidal
functors into a crafted monoidal target category; this was due to the lack of a
notion of \emph{displayed} monoidal categories that would have allowed one to
limit the ingredient to sections instead of those monoidal
functors. This displayed notion resisted definition efforts due to
the underlying notion of monoidal category that relies on a tensor
construed as a functor from $\CC\times\CC$ to $\CC$, with $\CC$ the
underlying category. For the displayed notion, it created major
difficulties with transport\footnote{Transport is used in univalent
  foundations to change a composite type of a term by applying an
  equality to a part of that type.} along components of pairs arising
with the use of this product category $\CC\times\CC$.
To solve this issue, instead of
working with two-place functions (encoded by pairing), we moved to a
curried view that first takes the left argument and then yields a function
that expects the right-hand side argument---which is good for the
object mapping. For the two-place morphism mapping, we instead employ
a symmetric approach, by considering the one-place mappings where the
left resp.\ right argument is fixed to the identity, which we call the
left resp.\ right \emph{whiskering}, respectively. The name
``whiskering'' comes from the analogous treatment of horizontal
composition in bicategories in \UniMath. In the end, monoidal
categories have become the one-object bicategories, but with a readable
syntax. This change opened the path to a usable definition of displayed monoidal
categories, fully solving the aforementioned problem.\footnote{To be
  found in the \UniMath GitHub repository by searching for the section
  names containing \texttt{FromMonoidalSection}.}

\section{Detecting Superfluous Assumptions}\label{sec:superfluous}
Here, we report on our experience with the formalization of
\cite[Theorem 4.7]{DBLP:conf/rta/FioreS17} by Fiore and Saville that was formulated as an
axiom in the formalization by Lafont of skewed monoidal
categories mentioned in Sect.~\ref{sec:monoidal}. The theorem is
announced as ``a conceptual framework for parametrised
iteratively-constructed initial algebras'' in the cited paper, and it
is instantiated to provide a proof of Lemma 4.8 there which says that
for a monoidal category $(\CC,I,\otimes)$ with some extra properties
and an $\omega$-cocontinuous strong bifunctor $F$ (from $\CC\times\CC$
to $\CC$), initial $F(Z,-)$-algebras are parametrised initial for
every object $Z$. The paper does not mention how to prove Theorem 4.7,
but we found out that it is an instance of the generalized Mendler
iteration scheme whose formalization Anders Mörtberg contributed to
\UniMath in 2016. However, the interesting element for our purposes
here is the presence of assumptions that are not needed for proving Lemma 4.8
and how we discovered that fact just by using
\Coq for the formalization: we formulated the relevant part of Theorem 4.7 with our notion
of whiskered bifunctors (Section~\ref{sec:bifunctor}) in a \Coq
section without proof, but with all the requirements with
\texttt{Context} commands. We then derived from it
Lemma 4.8, which was formulated similarly.
However, once
the proof of the theorem was given, its application for Lemma 4.8 did
no longer type-check, the simple reason being that \Coq had identified
unused assumptions in the proof of the theorem, which were hence no
longer abstracted when closing the \Coq section: they were $\omega$-cocontinuity of parameter $G$ and the condition that \A has $\omega$-colimits.
Also for Lemma 4.8,
the formalization suggested slightly weaker assumptions, but these are
the usual benefits of formalizing results.

\begin{acks}
  We gratefully acknowledge the work by the \Coq development team in providing the \Coq proof assistant and surrounding infrastructure, as well as their support in keeping \UniMath compatible with \Coq.
\end{acks}

\bibliographystyle{ACM-Reference-Format}
\bibliography{coqpl23abstract}

\end{document}